\documentclass[epj]{webofc}
\usepackage[varg]{txfonts}   

\usepackage{natbib}
\usepackage{csquotes}

\def\sun{\hbox{$\odot$}}

\newcommand\apj{\textit{ApJ}}%
\newcommand\apjl{\textit{ApJ}}%
\newcommand\aap{\textit{A\&A}}%
\newcommand\mnras{\textit{MNRAS}}%
\woctitle{Hot planets and cool stars}
\begin{document}
\title{The formation of planets by disc fragmentation}
%
%

\author{Dimitris Stamatellos\inst{1}\fnsep\thanks{\email{D.Stamatellos@astro.cf.ac.uk}}
}

\institute{School of Physics \& Astronomy, Cardiff University, Cardiff CF24 3AA, UK}

\abstract{%
I discuss the role that disc fragmentation plays in the formation of gas giant and terrestrial planets, and how this relates  to the formation of brown dwarfs and low-mass stars, and ultimately to  the process of star formation. Protostellar discs may fragment, if they are massive enough and can cool fast enough, but  most of the objects that form by fragmentation are brown dwarfs.  It may be possible that planets also form, if the mass growth of a proto-fragment  is stopped  (e.g. if this fragment is ejected from the disc), or suppressed and even reversed (e.g by tidal stripping).  I will discuss if it is possible to distinguish whether a planet has formed by disc fragmentation or core accretion, and mention of a few examples of observed exoplanets that are suggestive of  formation by disc fragmentation .}
\maketitle
\section{Introduction}
\label{intro}

There are two primary mechanisms for the formation of planets: (i) core accretion  \citep{Safronov:1969a,Goldreich:1973a,Mizuno:1980a,Bodenheimer:1986a,Pollack:1996a}, and (ii)~gravitational fragmentation of gas, i.e. in the same way that stars form.  Gas fragmentation produces stars and even substellar objects, i.e. objects with masses below the hydrogen-burning limit ($\sim 80~{\rm M}_{\rm J}$): brown dwarfs ($m\sim 13-80~{\rm M}_{\rm J}$; they are able to burn deuterium), and planets ($m <13~{\rm M}_{\rm J}$; they cannot sustain deuterium burning). There is no reason for gas fragmentation to stop either at the hydrogen-burning limit or the deuterium-burning limit: the minimum mass of an object that forms by gas fragmentation is given from the opacity limit for fragmentation which is thought to be $\sim1-5~{\rm M_{\rm J}}$ \citep{Low:1976a,Rees:1976a,Silk:1977a,Boss:1988a,Boyd:2005a,Whitworth:2006a,Boley:2010b,Kratter:2010b,Forgan:2011b,Rogers:2012a}. Substellar objects may form either from the collapse of very low-mass gravitationally-bound cores, which are produced by turbulent fragmentation of larger molecular clouds \citep{Padoan:2004a,Hennebelle:2008c}, or by  disc fragmentation  \citep{Kuiper:1951a,Cameron78,Boss:1997a, Whitworth:2006a, Stamatellos:2009a, Boley:2009a}.  The first mechanism may produce isolated/free- floating planets, whereas the later may produce both bound  and free-floating planets. 

It is generally accepted that rocky planets form via core accretion, but it still is debated how gas giant planets form.  In the core accretion model planets form by coagulation of dust particles in circumstellar discs. Some of these cores subsequently accrete envelopes of gas from the disc to become gas giants, whereas the rest end up as rocky planets.  In the disc fragmentation model gas giants form by direct gravitational fragmentation of protostellar gas in discs within a few thousand years.  However, in the last few years it has become evident that gravitational instabilities may play a more complex role in planet formation, probably working in tandem with core accretion, and contributing to the formation of  even rocky planets. 

\section{Observational evidence of disc fragmentation}
\label{sec-1}

The study of planet formation by gravitational instabilities in discs  has received new impetus by observations (i) of gas giant planets on wide orbits, and (ii) relatively large disc masses in the early stages of star formation. 

\paragraph{Giant planets on wide orbits and free-floating planets}

Gas giants on wide orbits are  difficult to form by core accretion \citep{Dodson-Robinson:2009a}. Such planets have been observed  by direct imaging, e.g. the four giant planets around HR8799 \citep{Marois:2008a,Marois:2010a}, and the giant planet around Fomalhaut \citep{Kalas:2008a}. Furthermore, statistical analysis of  microlensing surveys suggest that wide-orbit or free-floating  planets could be twice as common as main sequence stars \citep{Sumi:2011a}. 

\paragraph{Early stage massive discs}

Disc fragmentation happens fast, so that large, unstable discs or discs that are currently undergoing fragmentation should be rarely observed \citep{Stamatellos:2011d}.  Moreover, observations of early-stage discs are challenging as these discs are deeply embedded in their parental clouds and their presence has to be inferred from radiative transfer modelling. There have been a few candidates of  relatively massive discs in Class 0 and Class I objects \citep{Rodriguez05,Eisner:2005a,Eisner:2006a,Andrews07,Eisner:2008a,Jorgensen:2009a,Andrews:2009b,Isella:2009a}, in a few of which the disc status has been confirmed by (i) HCO$^+\ 3-2$ line observations revealing signs of Keplerian rotation \citep[e.g. L1489-IRS;][]{Brinch:2007a}, or (ii)  scattered light images \citep[L1527;][]{Tobin10}.   An interesting example of a massive disc is the one around HL~Tau (a 0.3-M$_{\sun}$ Class II object), which has a mass of around 0.1~M$_{\sun}$ and extends to at least 100 AU \citep{Greaves:2008a}. The frequency of occurrence of extended massive discs is uncertain as  a recent survey \cite{Maury:2010a} using the PdBI (1.3mm) found no discs larger than its resolution limits ($\sim100$~AU).

\section{Criteria for disc fragmentation}

 Gravitational instabilities grow in discs if two conditions are met: (i) gravity dominates over thermal and local centrifugal support \citep[the Toomre Q parameter is of order unity;][] {Safronov:1960a,Toomre:1964a}, and (ii) the disc cool fasts enough \citep[similar to or less than the local dynamical timescale;][]{Gammie:2001a,Johnson:2003a}.  For longer cooling times, the disc settles into a quasi-steady state in which angular momentum is transported outwards allowing mass to accrete onto the central star \citep[e.g.][]{Lodato:2004a,Lodato:2005b}. Recent work has shown that the inner disc region cools slowly compared to the local dynamical time; hence gravitational instabilities (inside $10 - 20$ AU) are typically suppressed \citep{Rafikov:2005a,Whitworth:2006a,Boley:2006a,Stamatellos:2008a}. Beyond  $\sim 70$ AU, these discs (if they have sufficient mass at these radii so that $Q\sim1$) cool efficiently and fragmentation becomes likely \citep{Whitworth:2006a,Stamatellos:2007b,Stamatellos:2009a,Boley:2009a}.

\section{Hydrodynamic simulations fragmenting discs}

\paragraph{Numerical approaches}

There have been two types of  numerical studies of disc fragmentation. In the first approach disc cooling is a free parameter \citep[e.g.][]{Gammie:2001a, Johnson:2003a,Rice:2005a,Clarke:2007a}; these studies have established that when the disc can cool faster than 0.5-2 orbital periods then it fragments. In the second approach, the disc thermodynamics are treated self-consistently in simulations with realistic equations of state that include the associated physics of molecular hydrogen and helium \citep[e.g.][see Figure~1]{Stamatellos:2007a,Boley:2007b}. The effects of the rotational and vibrational degrees of H$_2$, of H$_2$ dissociation, of H$^0$ ionisation, and  the first and second ionisation of helium are included. Moreover, opacity changes due to e.g. ice mantle melting, the sublimation of dust, molecular and H$^-$ contributions, are also taken into account.  

\paragraph{Disc fragmentation}

Numerical simulations   \citep{Stamatellos:2007a,Stamatellos:2009a,Stamatellos:2009d,Stamatellos:2011d} suggest that most of the objects that form by disc fragmentation are brown dwarfs ($\sim 80\%$) and low-mass stars  ($\sim 20\%$). The disc fragmentation model reproduces the brown dwarf desert, i.e. the lack of close companions to low-mass stars \cite[in contrast, planetary close companions and low-mass hydrogen burning star companions are frequently observed in this region;][]{Marcy:2000a, Grether:2006a, Sahlmann:2011a}. The objects that form in the disc initially  have masses of a few M$_{\rm J}$ but they grow by accreting material from the disc \cite{Stamatellos:2009d}. The objects that form first and migrate inwards gain enough mass to become stars, whereas the ones that stay in the outer disc region increase in mass but not as much, becoming brown dwarfs. If one of the brown dwarfs from the outer disc region drifts inwards, then it is quickly ejected again into the outer disc region due to  dynamical interactions with the higher-mass objects of the inner region.  The inner disc region will  also be populated by planets that form by core accretion at a later stage (after  $\sim1$~Myr). Most of the brown dwarfs are either ejected from the system becoming field brown dwarfs, or stay bound to the central star at relatively wide orbits ($\sim200-10^4$~AU). Therefore, it seems  that planets cannot form by disc fragmentation, unless the mass growth  of proto-fragments that form in the disc is  somehow suppressed.  
 
\begin{figure*}
\centerline{
\includegraphics[width=\textwidth]{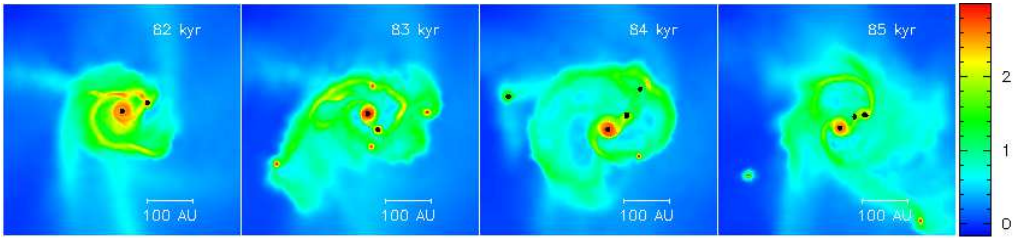}}
\caption{Evolution of the accretion disc around a protostar forming in a collapsing turbulent molecular cloud core. The disc increases in mass due to infalling material from the ambient cloud, becomes gravitationally unstable, and fragments.}
\label{fig:density}
\end{figure*}

 \paragraph{The formation of free-floating planets}
  
 One way to stop the mass growth of a proto-fragment that forms in the disc  is to  eject  it  from the disc quickly after its formation \cite{Stamatellos:2009a}. This will then become free-floating  and contribute to the population of free-floating planets \citep{Lucas:2000a, Osorio:2000a, Najita:2000a,Lodieu:2007a}, which microlensing observations suggest that it may be quite large \citep{Sumi:2011a}.

 \paragraph{Gas giant and rocky planet formation by tidal stripping}
 
Another way to suppress or even reverse the mass growth of proto-fragments forming in discs is by stripping away material through tidal effects as these fragments move closer to the central star \citep{Vorobyov:2006a, Boley:2010b,Nayakshin:2010a,Vorobyov:2010a,Boss:2012a, Liu:2013a}.
 In this model proto-fragments form at distances $\stackrel{>}{_\sim}70$~AU and start migrating inwards, while contracting on a Kelvin-Helmholtz time scale, i.e. relatively slowly. During this phase, grains can sediment to the centre of the fragment forming a solid core. When the proto-fragment approaches the
central star tidal forces strip material from the envelope of  this core and the mass of the proto-fragment reduces; it may even become a rocky planet \citep{Vorobyov:2006a, Boley:2010b,Nayakshin:2010a,Vorobyov:2010a,Boss:2012a}.

\paragraph{Discs forming self-consistently in collapsing molecular clouds}

Numerical models have also studied discs that self-consistently form in collapsing clouds. Such studies  have demonstrated the importance of both disc-core interaction and variable accretion in driving repetitive episodes of disc fragmentation \citep{Vorobyov:2005a,Kratter:2010a,Vorobyov:2010a, Vorobyov:2010e, Machida:2010a, Rice:2010a,Stamatellos:2011c,Zhu:2012a, Stamatellos:2012a}.  Mass infall can have a large impact on the stability of the outer disc and provides a physical mechanism for driving the system to an unstable state.

\paragraph {Planet formation in clusters}

Most stars form in clusters and therefore discs form and evolve in an environment where interactions may be frequent \citep{McDonald:1995a,Thies:2005a}. Numerical studies including realistic disc heating/cooling suggest that perturbations from passing stars result  in tidal heating  that may suppress disc fragmentation  in the inner disc region \citep{Lodato:2007a, Forgan:2009d}. On the other hand, it is possible that disc instabilities may be promoted on the outer regions of extensive discs \citep{Thies:2010a,Shen:2010a}.

\paragraph {Planet formation in binaries} 

A large fraction of of stars are in close or wide binaries, where the components of the binary will affect the evolution of circumstellar and circumbinary discs. Currently there is no consensus on whether such interactions promote or suppress disc fragmentation \citep{Mayer10}. Disc fragmentation and planet formation may be promoted in circumbinary discs as matter accumulates in a ring around the central stars and spiral waves are excited and propagate in the disc (Stamatellos et al. in prep).

\section{Can we distinguish between planets formed by disc fragmentation and core accretion?}

\paragraph{Mass}
It was thought that terrestrial planets are formed by core accretion and gas giant either by core accretion or gas fragmentation, but recent ideas \cite[e.g. tidal stripping/downsizing;][]{Boley:2010b,Nayakshin:2010a, Liu:2013a} are challenging this view.

\paragraph{Orbital radius} 
It is expected that disc fragmentation will produce planets on wide orbits,  and core accretion on close orbits to the central star, but it is rather unlikely that these planets will remain where they form. Dynamical interactions may either scatter planets formed close to the star (by core accretion) outwards  \citep{Veras12} or scatter planets (formed by disc fragmentation on wide orbits) inwards \citep[e.g.][]{Stamatellos:2009a}. Moreover, interactions of proto-fragments with the disc may cause inward migration \citep{Vorobyov:2006a, Stamatellos:2009d,Boley:2010b,Ayliffe:2010a,Ayliffe:2011a, Michael:2011a,Baruteau:2011a}. 

\paragraph{Metallicity}
I may be argued that planets that form by gravitational instabilities have the same metallicity as the host star (i.e. the metallicity of the cloud collapsing to form that star, presumably the metallicity of the ISM), whereas objects formed by core accretion have enhanced metallicities due to the higher proportion of dust grains going into planetesimals. However, recent work suggests that   gas giants that form by disc fragmentation may have a variety of heavy-element compositions, ranging from sub- to super-nebular values. High levels of enrichment can be achieved through  enrichment at birth, planetesimal capture, and differentiation plus tidal stripping \citep{Helled:2006a,Boley:2011a, Helled:2010a}.
 
 \paragraph{Age}
 
 Observing young planets around extremely young stars may point towards formation by disc fragmentation e.g the candidate proto-planet around  HL Tau \citep{Greaves:2008a}. The central star  is thought to be only  $\sim0.1$-Myr old, and considering the large mass of the suspected proto-planet ($\sim 14~{\rm M}_{\rm J}$) it is rather unlikely that it have formed by core accretion within such a short time.

\section{Exoplanets formed by disc fragmentation?}

There is a growing number of observed exoplanets, whose mass and distance from the central star, pose a problem to the core accretion scenario, and therefore are suggestive of formation by disc fragmentation. {\bf (i)} HL Tau: a candidate proto-planet  ($\sim 14~{\rm M_J}$) on a wide orbit ($\sim 65$~AU) embedded in a $\sim 0.1-{\rm M_{\sun}}$ disc around a 0.3-M$_{\sun}$ star \citep{Greaves:2008a}. {\bf (ii)}~HR8799: a 4-planet system with four giant planets (each one with mass $\sim 10~{\rm M_J}$) on wide orbits (15-70~AU) around a 1.5-M$_{\sun}$ A-type star \citep{Marois:2008a,Marois:2010a}. {\bf (iii)}~2MASS1207: a planet  ($\sim 5-8~{\rm M_J}$)  at distance $\sim 55$ AU from the 25~${\rm M_J}$ brown dwarf primary \citep{Chauvin04,Chauvin05}. {\bf (iv)} Fomalhaut: a candidate planet  ($\sim 3~{\rm M_J}$)  at distance $\sim 115$ AU around a 2-M$_{\sun}$ star \citep{Kalas:2008a}. {\bf (v)}~1RXS J160929.1-210524: a planet  ($\sim 6-11~{\rm M_J}$)  at distance $\sim 330$ AU from a 0.7-M$_{\sun}$ star \citep{Lafreniere:2008a,Lafreniere:2010a}.

\section{Conclusions}

It is possible that protostellar discs may fragment and produce low-mass mass objects; these may be planets, brown dwarfs, and/or low-mass hydrogen-burning stars. Numerical simulations  of disc fragmentation which have followed the evolution of  proto-fragments  in discs suggest the following outcomes: 
\begin{itemize}
\item The proto-fragments formed by fragmentation survive but their growth is suppressed or even reversed (e.g. by tidal effects) and end up as gas giant or rocky planets \citep{Vorobyov:2006a, Boley:2010b,Nayakshin:2010a,Vorobyov:2010a,Boss:2012a}, on close or wide orbits depending on the effects of migration. 
\item The proto-fragments survive but  they grow in mass: proto-fragments that migrate inwards accrete more and become low-mass stars, whereas those that stay on wide orbits become brown dwarfs \citep{Stamatellos:2007a,Stamatellos:2009a,Stamatellos:2009d}. 
\item The proto-fragments survive but they are ejected from the disc ending up as field objects, i. e. free floating planets, brown dwarfs and low-mass stars, depending on their mass they manage to accrete before ejection \citep{Stamatellos:2009a,Stamatellos:2009d,Kratter:2010a,  Kratter:2011b, Basu:2012a}.
\item The proto-fragments are disrupted and die out.
\end{itemize}

Therefore, it is possible that some planets may form disc fragmentation. Even if gravitational instabilities and disc fragmentation do not lead to the formation of planets except in rare circumstances, they may play a fundamental role in promoting planetesimal formation and growth \citep{Durisen:2005a,Vorobyov:2005a,Rice:2006a,Rice:2006b,Vorobyov:2006a,Clarke:2009b,Vorobyov:2010c,Machida:2011a}. 

\begin{acknowledgement}
I would like to thank Ant Whitworth, Aaron Boley,  Shu-ichiro Inutsuka, Kaitlin Kratter, Ken Rice, Eduard Vorobyov,  and Doug Lin, for stimulating discussions  about forming planets and brown dwarfs by disc fragmentation. I also thank the organisers of the conference that gave me the opportunity to meet and interact with the researchers of the RoPACS network.
\end{acknowledgement}


\end{document}